\documentclass[aps,preprint,preprintnumbers,amsmath,amssymb,amsfonts,groupedaddress,superscriptaddress,showpacs,showkeys,floatfix]{revtex4-1}

\usepackage{graphicx}
\usepackage{subfigure}
\usepackage{bm}
\usepackage{color}
\usepackage{array}
\usepackage{epstopdf}

\begin{document}

\title{Possible strain induced Mott gap collapse in 1\emph{T}-TaS$_2$}

\author{Kunliang Bu}
\affiliation{Zhejiang Province Key Laboratory of Quantum Technology and Device, Department of Physics, Zhejiang University, Hangzhou 310027, China}
\author{Wenhao Zhang}
\affiliation{Zhejiang Province Key Laboratory of Quantum Technology and Device, Department of Physics, Zhejiang University, Hangzhou 310027, China}
\author{Ying Fei}
\affiliation{Zhejiang Province Key Laboratory of Quantum Technology and Device, Department of Physics, Zhejiang University, Hangzhou 310027, China}
\author{Zongxiu Wu}
\affiliation{Zhejiang Province Key Laboratory of Quantum Technology and Device, Department of Physics, Zhejiang University, Hangzhou 310027, China}
\author{Yuan Zheng}
\affiliation{Zhejiang Province Key Laboratory of Quantum Technology and Device, Department of Physics, Zhejiang University, Hangzhou 310027, China}
\author{Jingjing Gao}
\affiliation{Key Laboratory of Materials Physics, Institute of Solid State Physics, Chinese Academy of Sciences, Hefei 230031, China}
\affiliation{University of Science and Technology of China, Hefei 230026, China}
\author{Xuan Luo}
\affiliation{Key Laboratory of Materials Physics, Institute of Solid State Physics, Chinese Academy of Sciences, Hefei 230031, China}
\author{Yu-Ping Sun}
\affiliation{Key Laboratory of Materials Physics, Institute of Solid State Physics, Chinese Academy of Sciences, Hefei 230031, China}
\affiliation{High Magnetic Field Laboratory, Chinese Academy of Sciences, Hefei 230031, China}
\affiliation{Collaborative Innovation Center of Advanced Microstructures, Nanjing University, Nanjing 210093, China}
\author{Yi Yin}
\email{yiyin@zju.edu.cn}
\affiliation{Zhejiang Province Key Laboratory of Quantum Technology and Device, Department of Physics, Zhejiang University, Hangzhou 310027, China}
\affiliation{Collaborative Innovation Center of Advanced Microstructures, Nanjing University, Nanjing 210093, China}

\begin{abstract}
Tuning the electronic properties of a matter is of fundamental interest in scientific research as well as in applications.
Recently, the Mott insulator-metal transition has been reported in a pristine layered
transition metal dichalcogenides 1\emph{T}-TaS$_2$, with the transition
triggered by an optical excitation, a gate controlled intercalation, or a voltage pulse.
However, the sudden insulator-metal transition hinders an exploration of how the transition evolves.
Here, we report the strain as a possible new tuning parameter to induce Mott gap collapse in 1\emph{T}-TaS$_2$.
In a strain-rich area, we find a mosaic state with distinct electronic density of states within different domains.
In a corrugated surface, we further observe and analyze a smooth evolution from a Mott gap state to a metallic state.
Our results shed new lights on the understanding of the insulator-metal transition and promote a controllable
strain engineering on the design of switching devices in the future.
\end{abstract}

\maketitle

For a half-filled electronic band, strong correlation of electrons can
lead to a unique Mott insulator state, when the ratio of Coulomb repulsion
$U$ to the bandwidth $W$ ($U/W$) exceeds a critical value~\cite{MottRMP68}.
Proximity to the Mott insulator is the origin of many exotic superconducting states,
such as in cuprates~\cite{LeeRMP06}, magic-angle
graphenes~\cite{CaoNAT18, CaoNAT18(1)} and transition metal
dichalcogenides~\cite{SiposNMAT08, LiEPL12, LiuAPL13}. To explore the superconducting mechanism,
it is important to understand the Mott insulator state and how the transition evolves
from a Mott insulator to a metallic or superconducting state~\cite{CaiNPHYS16}.

The transition metal dichalcogenide 1\emph{T}-TaS$_2$ is a correlation-induced
Mott insulator~\cite{SiposNMAT08, LiEPL12, LiuAPL13}. The Mott insulator state of 1\emph{T}-TaS$_2$ is not residing
in a periodic atomic lattice, but in a periodic commensurate charge density wave (CCDW) lattice.
With the CDW state susceptible to external perturbations, the Mott insulator
state is relatively easy to be modulated~\cite{YuNNANO15, TsenPNAS15}.
The Mott phase of the pristine
1\emph{T}-TaS$_2$ can be suppressed by chemical
doping~\cite{AngPRL12, AngPRB13, AngNATCOMMUN15, QiaoPRX17, FujisawaJPSJ17}, intercalation~\cite{LahoudPRL14, WagnerPRB08},
thermal excitation~\cite{SalvoPRB75}, or pressure~\cite{SiposNMAT08}.
Recently, the Mott insulator-metal transition has been further controlled by
an optical excitation~\cite{StojchevskaSCI14, VaskivskyiSCIADV15},
gate controlled intercalation~\cite{YuNNANO15}, or a charge
pulse~\cite{HollanderNANOLETT15, VaskivskyiSCIADV15, VaskivskyiNATCOMMUN16, ChoNATCOMMUN16, MaNATCOMMUN16, YoshidaSCIADV15}.
Although the underlying mechanism of insulator-metal transition is still under debate,
these controlled tuning methods are good candidates for the design of switching devices.
Insightful information has been reported to tune the electronic
states of complex materials by strain~\cite{ZhaoNANOLETT17, MeyersPRB13, TorreNATCOMMUN15, AhnSCI04, HanNANOLETT18, DadgarCM18, AndradearXiv18, GanPCCP16,GaoPNAS18,SvetinAPE14}.

In this study, we show a possible strain-induced Mott-gap collapse in the pristine 1\emph{T}-TaS$_2$ by scanning tunneling microscopy (STM). In a strain-rich area,
we find mosaic CDW domains and the stable mosaic state is most possibly induced
by the intrinsic strain. In the mosaic state, we could detect variable spectra
from a Mott gap state to a metallic state within different domains.
We further find a corrugated surface, also possibly a strain-induced feature.
When being across the corrugation, a smooth evolution of the Mott-gap collapse
is observed and analyzed. The Mott gap is suppressed gradually and a V-shaped metallic state
emerges at the corrugation. In the process of Mott-gap
collapse, the rapid increase of bandwidth $W$ is found to be the dominant
factor to reduce $U/W$. By gluing 1\emph{T}-TaS$_2$ on the organic-glass substrate, we introduce strain to the sample surface at low temperature,
and confirm the strain-induced mosaic pattern and corrugation. Our results provide a further understanding of the Mott
insulator-metal transition and suggest the strain engineering as a possible new tuning
method to modulate the Mott insulator state.

\noindent\textbf{Results}

\begin{figure}
\centering
\includegraphics[width=.8\columnwidth]{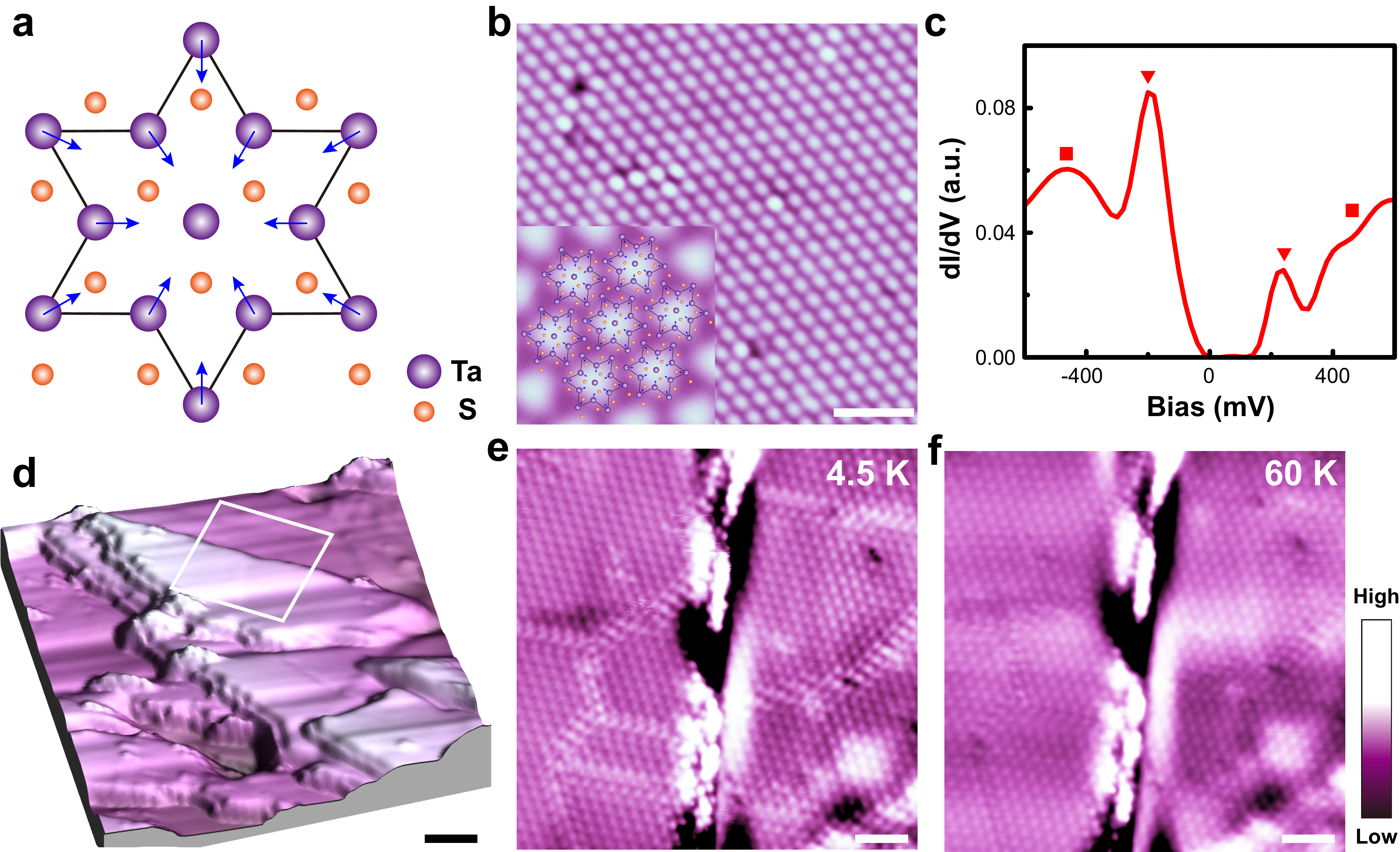}
\caption{Stable mosaic state in 1\emph{T}-TaS$_2$. \textbf{a} Schematic
structure of a Star of David (SOD). The purple and yellow balls represent the Ta and S atoms, respectively.
The blue arrows indicate that the 12 surrounding Ta atoms shrink to the central Ta atom.
\textbf{b} A typical topography of 25 nm $\times$ 25 nm in a clean area. Inset shows
a 4 nm $\times$ 4 nm image with the schematic structure of SODs superimposed on it. The tunneling
condition is $V_{\mathrm b} = 600$ mV and $I = 100$ pA. \textbf{c} Average $dI/dV$ spectra
simultaneously taken with \textbf{b}. The bias modulation is set to be 10 mV. The inverted
triangles and squares indicate coherence peaks of Mott gap and charge density wave (CDW) gap, respectively.
\textbf{d} A three dimensional plot of the topography over a 100 nm $\times$ 100 nm
area. The tunneling condition is $V_{\mathrm b} = 1$ V and $I = 20$ pA.
\textbf{e, f} The enlargement of the white box (38 nm $\times$ 38 nm) at 4.5 K (\textbf{e})
and at 60 K (\textbf{f}). The tunneling condition is $V_{\mathrm b} = 600$ mV,
$I = 300$ pA in \textbf{e}, and $V_{\mathrm b} = 1$ V, $I = 200$ pA in \textbf{f}.
Scale bar is 10 nm in \textbf{d} and 5 nm in \textbf{b}, \textbf{e}, and \textbf{f}.
}
\label{fig_n01}
\end{figure}

\noindent\textbf{Crystal structure and electronic state of the pristine 1\emph{T}-TaS$_2$.}
The unit structure of 1\emph{T}-TaS$_2$ is composed of
a triangular lattice of Ta atoms sandwiched between two layers of triangular lattice of
S atoms. With an ABC-type stacking, each Ta atom is coordinated octahedrally by S atoms.
The sample is at near commensurate CDW (NCCDW) state at
room temperature and develops to the CCDW state at around 170 K (Supplementary Fig. 1).
A basic element of the CDW state is the so called Star of David (SOD). As shown in Fig.~\ref{fig_n01}a,
The SOD cluster is formed by 13 Ta atoms with 12 surrounding Ta atoms shrinking to the central Ta.
At the CCDW state, the SODs are regularly arranged to form a $\sqrt{13} \times \sqrt{13}$
reconstruction~\cite{WilsonADVPHYS75, FazekasPMB79, FazekasPBC80}. With the single crystal
sample cleaved and a top S layer exposed, the
STM experiment is performed on the exposed surface at liquid helium temperature around $4.5$ K.

A typical topography is shown in Fig.~\ref{fig_n01}b, with each bright spot representing a SOD.
Consistent with the $\sqrt{13} \times \sqrt{13}$ reconstruction, the bright spots form a
triangular lattice of SODs and the distance of two neighboring SODs is
measured to be 12.1 \AA. A single unpaired $5d$ electron of the central Ta atom in each
SOD contributes to the half-filled electronic band, leading to the Mott insulator
state of the pristine 1\emph{T}-TaS$_2$. In the field of view (FOV) in Fig.~\ref{fig_n01}b,
a differential conductance ($dI/dV$) spectrum is measured at each point in a dense grid of
spatial positions. The average spectrum is shown in Fig.~\ref{fig_n01}c.
With the spectrum proportional to the electronic density of
states (DOS), the sharp coherent peaks at 240 mV and -200 mV correspond to the upper
and lower Hubbard bands (UHB and LHB), respectively~\cite{AngPRL12}. Energy positions of UHB and LHB
result in a Mott gap of 440 $\pm$ 20 meV. In the average spectrum, there is also
a broad peak at -460 mV  and a kink feature at 440 mV, corresponding to the valence band (VB)
peak and the conduction band (CB) peak of the CDW gap, respectively~\cite{AngPRL12, ChoPRB15}.
All these characteristics are consistent with previous reports of the pristine
1\emph{T}-TaS$_2$~\cite{ChoPRB15, ChoNATCOMMUN16, ChoNATCOMMUN17, MaNATCOMMUN16, QiaoPRX17}.

\noindent\textbf{Stable mosaic state in a strain-rich area.}
We intentionally look for a strain-rich area.
Figure~\ref{fig_n01}d is a topographic image
of a 100 nm $\times$ 100 nm area. The complex morphology indicates a strain-rich
environment around this area, which may originate from the cleavage process.
A zoom-in image of
the white box shows a mosaic state with several nanometer-sized domains (Fig.~\ref{fig_n01}e).
The textured domains are different from the quasi-hexagonal phase in
the NCCDW state~\cite{WuPRL90, SpijkermanPRB97}. The pattern shows that
they are more similar to the voltage pulse induced mosaic
state~\cite{MaNATCOMMUN16, ChoNATCOMMUN16}. Within each domain, the superlattice of SODs
is still preserved. Neighboring domains are separated by bright domain walls,
across which there is a translational phase shift of the CDW order. We do not
see any rotational shift of the CDW order between different domains~\cite{QiaoPRX17}.
This mosaic state is stable
at 4.5 K, without any change after a longtime measurement. With the temperature increased to
60 K (Fig.~\ref{fig_n01}f), the domain wall pattern is almost the same as
that at low temperature (see more details in Supplementary Fig. 2 and Supplementary Fig. 3).
Different from the metastable mosaic state triggered by a voltage pulse~\cite{ChoNATCOMMUN16, MaNATCOMMUN16},
this mosaic state in strain-rich area is very stable, possibly attributed to the intrinsic and stable strain.

\begin{figure}[tp]
\includegraphics[width=.95\columnwidth]{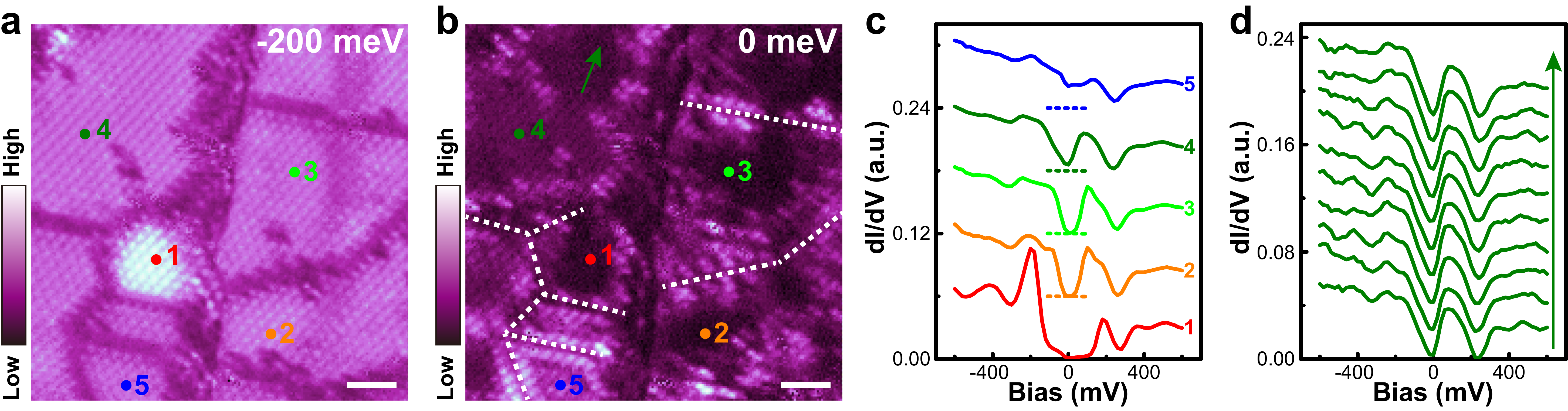}
\caption{Electronic states of mosaic domains. \textbf{a, b} The differential
conductance maps at -200 mV (the energy of lower Hubbard band) and 0 mV (Fermi energy), respectively.
Scale bar is 5 nm.
The field of view is approximately the same as that in Figs. 1\textbf{e} and 1\textbf{f}.
The white dashed lines in \textbf{b} illustrate the domain walls. \textbf{c} The $dI/dV$
spectra at marked positions in \textbf{a} and \textbf{b}.
The dataset is shifted vertically for clarity. The short dashed lines represent the zero
vertical coordinate of shifted curves.
\textbf{d} A series of $dI/dV$ spectra taken along the olive arrow in \textbf{b}.
All spectra are taken at $V_{\mathrm b} = 600$ mV and $I = 100$ pA, with a bias modulation of 10 mV.
}
\label{fig_n02}
\end{figure}

\noindent\textbf{Electronic states in mosaic domains.}
Figures~\ref{fig_n02}a and
\ref{fig_n02}b show conductance maps of the mosaic domains under a bias
voltage of -200 mV and 0 mV, respectively. The bias voltage of -200 mV
is chosen to be at the peak position of LHB.
The pattern of mosaic domains can be clearly observed in Fig.~\ref{fig_n02}a,
with the domain walls one-to-one mapped to Fig.~\ref{fig_n02}b (white dashed lines).
Five typical $dI/dV$ spectra are taken within different domains,
at labeled positions (Fig.~\ref{fig_n02}c). For the position marked by `1',
the $dI/dV$ spectrum shows a Mott insulator state, with
sharp coherence peaks of LHB and UHB. For spectra at positions marked from `2' to `5', the differential
conductance at -200 mV decreases, the Mott gap is suppressed to a V-shaped gap,
and the V-shaped gap develops together with a finite DOS at zero bias (Fermi energy $E_\textrm{F}$).
A V-shaped gap has been observed in Cu intercalated 1\emph{T}-TaS$_2$,
and the gap gradually disappears with the increase of temperature~\cite{LahoudPRL14}.
A similar V-shaped gap has also been observed in isovalent Se doped 1\emph{T}-TaS$_2$~\cite{QiaoPRX17}.

The variation of $dI/dV$ spectra is consistent with the conductance map in Fig.~\ref{fig_n02}a,
in which the domain marked by `1' is represented by a bright white patch and different from
other purple patches. We notice that the conductance is relatively uniform within each domain in Fig.~\ref{fig_n02}a,
and the periodic pattern in the domain is still consistent with
the CCDW superlattice. For conductance at $E_\textrm{F}$, both a zero conductance
and a finite conductance are observed in the $dI/dV$ spectra in Fig.~\ref{fig_n02}c, with the latter
representing a metallic state (olive and blue curves).
Within each single domain,
the zero bias conductance at the clean area is also homogeneous, as shown in Fig.~\ref{fig_n02}d.
Some of the bright features in Fig.~\ref{fig_n02}b are due to the CDW impurities like missing or distorted SOD.
Other bright patches in Fig. 2b aggregate at the step edge and the edge of each domain. There
may be a mechanism that leads to trapped carriers at the edges of domains, which is however not clear yet.

\noindent\textbf{Mott-gap collapse at a corrugation.}
The mosaic domains and related complex electronic states are speculated to be induced by the
intrinsic strain.
We further find an area with a corrugated surface,
not far away from the strain-rich area (Supplementary Fig. 4). The corrugation is also
a possible strain-induced feature~\cite{OkadaNATCOMMUN12, XieSCI18}.
In this corrugated area, we observe a smooth evolution of
Mott-gap collapse across the corrugation, which gives us a special example to analyze
the Mott insulator-metal transition.

\begin{figure}[tp]
\includegraphics[width=.8\columnwidth]{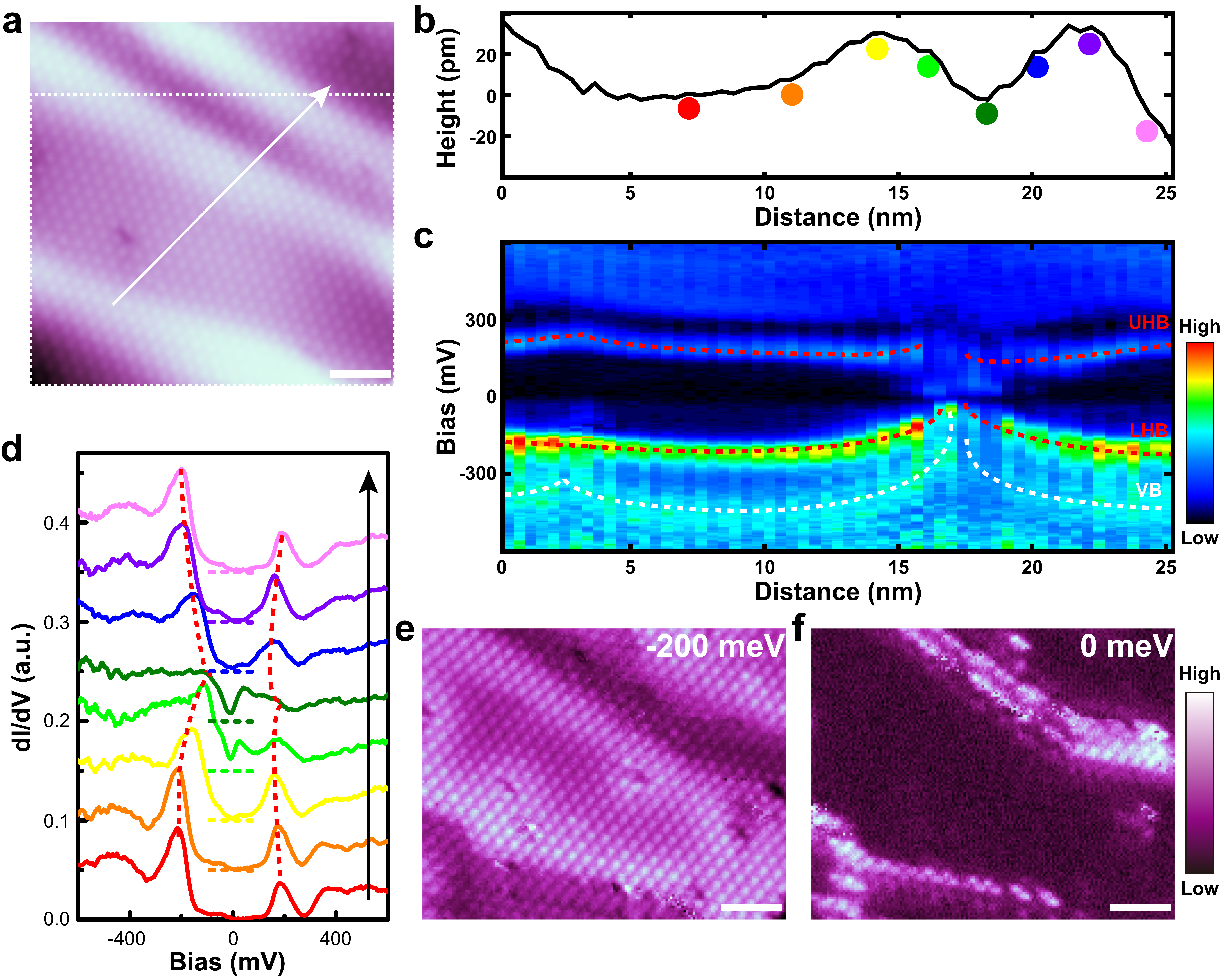}
\caption{Mott-gap collapse across the corrugation. \textbf{a} A topographic image with
corrugations (30 nm $\times$ 30 nm). The tunneling condition is $V_{\mathrm b} = 1$ V
and $I = 200$ pA. \textbf{b} The height profile along the white arrowed line in \textbf{a}.
The height profile is obtained from collected tip height value in the topography, with a linear
background subtraction. \textbf{c} The acquired $dI/dV$ spectra along the height profile, as
a function of the distance along the arrowed line. The red and white dashed lines guide
eyes to the Hubbard bands and valence band, respectively. The spectra are taken with a
tunneling condition of $V_{\mathrm b} = 600$ mV, $I = 100$ pA and a bias modulation of 10 mV.
\textbf{d} The $dI/dV$ spectra corresponding to the positions marked by colored dots in \textbf{b}.
The red dashed lines guide eyes to the evolution of
Hubbard bands.  \textbf{e, f} Differential conductance maps at -200 mV and the Fermi level,
respectively. The field of view is the same as the white dashed box in \textbf{a}. The conductance map is
taken at $V_{\mathrm b} = 600$ mV and $I = 100$ pA, with a bias modulation of 10 mV.
Scale bar in \textbf{a}, \textbf{e}, and \textbf{f} is 5 nm.
}
\label{fig_n03}
\end{figure}

Figure~\ref{fig_n03}a shows a topographic image of the corrugated surface, from which
we could observe a periodic and triangular lattice of SODs without any domain walls.
This FOV is within a single domain. Bright and dark stripes appear roughly
along the diagonal direction, representing a corrugated surface with a modulation of $z$-axis height.
We focus on a straight line across central stripes (the white arrowed line in Fig.~\ref{fig_n03}a).
In Fig.~\ref{fig_n03}b, the surface height
is drawn as a function of distance along the line, in which the ups and downs
correspond to the bright and dark stripes in the two-dimensional topograph.
A flat surface is shaped into a corrugation with parallel ridges and grooves.
The height of the corrugation is in the range of tens of picometers, comparably small
as that of other stain-induced corrugations~\cite{SlezakPNAS08, OkadaNATCOMMUN12, ZeljkovicNNANO15}.
Without an atomic resolution in this experiment, we cannot make a quantitative analysis
of the strain based on precise determination of atomic displacement~\cite{GaoPNAS18, ZeljkovicNNANO15}.

We measure a series of $dI/dV$ spectra along the straight line, with data shown in Fig.~\ref{fig_n03}c.
Some typical spectra are selectively chosen and shown in
Fig.~\ref{fig_n03}d. The location of each spectrum is labeled by a colored dot in the height
profile (Fig.~\ref{fig_n03}b). Approaching the dark groove from both sides,
we could observe a smooth evolution of Mott-gap collapse. Both UHB and LHB peaks move
gradually toward the zero bias (Fermi energy $E_\textrm{F}$), with peak height decreases and peak
width increases. The energy range of zero conductance shrinks until an in-gap state develops
to form a metallic V-shaped spectrum. From the smooth evolution of spectra, we can
track how the Hubbard band peaks evolve when approaching the groove.
In the metallic state, the Hubbard band peaks are separated from two V-spectrum peaks.
In Fig.~\ref{fig_n03}c, another important feature is that the VB peak moves toward the zero bias
together with the LHB peak, and finally merges to form the V-shaped spectrum.

To check the detailed distribution of the Mott-gap collapse in this corrugated area, we choose
a framed area in Fig.~\ref{fig_n03}a and measure the $dI/dV$ spectra at a dense array of locations.
Figures~\ref{fig_n03}e and~\ref{fig_n03}f show conductance maps at -200 mV and 0 mV, respectively.
Consistent with previous linecut spectra, metallic state in the shallow groove corresponds
to a dark depression in Fig.~\ref{fig_n03}e and a bright protrusion in Fig.~\ref{fig_n03}f.
In contrast, the Mott insulator state outside the groove shows a strong LHB peak (bright color in
Fig.~\ref{fig_n03}e) and a zero conductance around the Fermi energy (dark color in Fig.~\ref{fig_n03}f).
An additional metallic state at the left bottom corner in Fig.~\ref{fig_n03}f may be due to the complex trough there,
as indicated by the large depression at the same position in Fig.~\ref{fig_n03}a.

\noindent\textbf{Discussion}

\begin{figure}[tp]
\includegraphics[width=0.5\columnwidth]{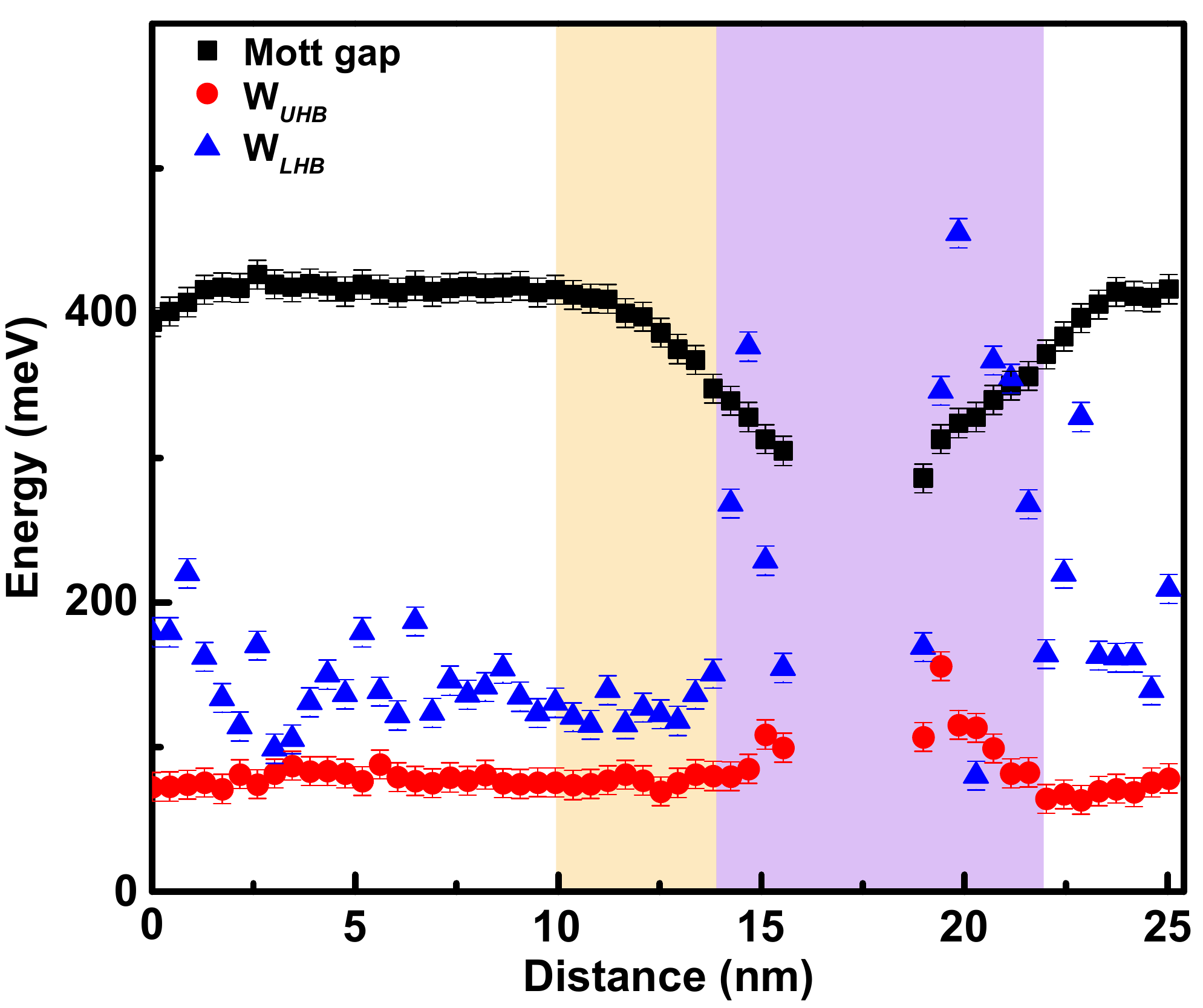}
\caption{Evolution of the electronic states across the corrugation. The $dI/dV$ spectra
in Fig.~\ref{fig_n03}\textbf{b} are fitted by a Gaussian function, to extract the Mott gap, bandwidth of
upper Hubbard band (UHB) and bandwidth of lower Hubbard band (LHB) (represented by black squares, red dots and blue triangles, respectively).
The data in the vicinity of the groove cannot be fitted reasonably, correspondingly not shown.
The error bars for the measured data points are due to the bias modulation.
Within the brown region, the Mott gap is gradually reduced and the bandwidth is nearly unchanged.
For the purple region, the Mott gap is reduced together with a rapid increase of the bandwidth of LHB.
}
\label{fig_n04}
\end{figure}

A Mott insulator-metal transition can be generally explained by a reduced ratio of $U/W$ in the
one-band Hubbard model.
The change of $U$ can be represented
by a change of the Mott gap. For the linecut spectra in Fig.~\ref{fig_n03}c, we fit both UHB and LHB
peaks of each spectrum with a Gaussian function (Supplementary Fig. 5)
and extract their energy positions and bandwidths. As shown in Fig.~\ref{fig_n04},
the bandwidth of LHB is consistently larger than that of UHB,
indicating a higher hopping constant $t$ for LHB ($t$ is proportional to $W$).
When approaching the groove from the left side, we could observe
that the Mott gap gradually decreases while the bandwidth of both UHB
and LHB peaks is nearly unchanged (brown region).
Here the decrease of $U$ is mainly responsible for the reduced $U/W$. When being closer to the
groove (purple region), the VB gradually merges with LHB and the LHB bandwidth
rapidly increases. The increase of UHB bandwidth is accompanied by the development of
in-gap state and the V-shaped spectrum. Although $U$ still decreases in this region, the
increase of bandwidth is observed to be the dominant factor to reduce $U/W$.

In Ref.~\cite{QiaoPRX17}, a multi-orbital Hubbard model has been proposed, including
contributions both from the central Ta orbital and the edge orbital of surrounding Ta atoms.
With the one-band Hubbard model as a down-folded version of the multi-orbital model, the reduction
of $U/W$ is explained to be both from the decrease of $U$ (of central
Ta orbital) and from the decrease of on-site energy difference between two orbitals.
The change of $U$ can be represented by a change of the Mott gap, and the
on-site energy difference between two orbitals can be represented by the relative position
between VB and LHB peaks. Consistent with this two-orbital model, our observation
reveals that the rapid increase of $W$ happens together with the mergence between
VB and LHB, which is the main factor to reduce $U/W$ and induce a metallic state.
The smooth spectrum evolution in our corrugated surface provides a clear picture
of how the Mott insulator-metal transition gradually happens. A Mott-gap collapse
could also happen on the intrinsic domain walls~\cite{FujiiJPCS18, ChoNATCOMMUN17, SkolimowskiPRL19},
which is however a sharp transition that we cannot obtain a similar analysis as
in Fig.~\ref{fig_n04} (see Supplementary Note 1). The single domain in the
topographic image of the corrugated surface also proves that this Mott-gap
collapse is a new phenomenon different from that induced by the domain-wall.

We also find another corrugation at this sample,
which shows similar Mott-gap collapse at the corrugation (Supplementary Fig. 6).
The mergence of the VB and LHB is reproduced at this corrugation.
We further conducted experiment on the `strained sample' by
gluing 1\emph{T}-TaS$_2$ on the organic glass substrate.
With a large thermal expansion coefficient of the substrate,
a compressive strain is expected to act on the sample at low temperatures.
The mosaic state and the corrugation are both confirmed in the `strained sample' (Supplementary Fig. 7 and Supplementary Fig. 8).
We also conducted the experiment on samples glued on the SiO$_2$ (0001) substrate,
which has a rather small thermal expansion coefficient and is expected to bring a tensile
strain to the glued sample. Instead of corrugations, cracks are the dominant features
on the surface for samples glued on the SiO$_2$ (0001) substrate.
The precise position of the Mott-gap collapse cannot be simply concluded from the
height profile of corrugations. The atomic-resolved topography is required
for the microscopic analysis of the strain distribution in a corrugation,
which is however technically challenging. Therefore to answer how or whether
the strain as an underlying mechanism tunes the Mott-gap collapse is still beyond our current
capabilities. Nevertheless, the observation of the similar behavior in the `strained sample' suggests that
the stable mosaic state and the corrugation can be generated as a response to the strain.

The mosaic state is a very complex phenomenon. Following we discuss possible roles
of strain for the mosaic state. For the mosaic state, an energy cost is
required to create domain walls. Strain may help to overcome the energy barrier for the creation of domain walls.
Then the mosaic state can be initially generated as a response to the strain.
Other techniques like laser pulse or tip-current pulse can similarly help to
create domain walls and mosaic state in this material. No matter how the mosaic
pattern is created, the metallic state in mosaic pattern can be similarly
related with the destruction of the long range CCDW order. Afterwards, a
global strain may possibly make the domain pattern energetically stable.
A strain can tune the electronic states while preserves the overall flat
atomic plane~\cite{GaoPNAS18}, like the rather weak spatial height modulation
within the mosaic state. Another proposal to explain the mosaic state
is that the stacking order tunes the Mott insulator-metal
transition~\cite{ChoNATCOMMUN16, MaNATCOMMUN16}. The stacking order is variable
according to the SOD misalignment along the
crystalline $c$ axis~\cite{RitschelNPHYS15, HovdenPNAS16}.
In this experiment, we did not observe any domain walls underneath the top layer, thus cannot provide
evidence for the stacking order proposal. A single uniform domain of Mott
insulator state exists in the strain-rich area (Fig.~\ref{fig_n02}), which
may be related with the stacking order effect.

In conclusion, we have carefully studied the strain-induced corrugation and mosaic state in 1\emph{T}-TaS$_2$.
The smooth evolution of spectrum
in corrugated surface provides a clear picture of how the Mott gap collapses.
The mosaic state and the corrugation is further confirmed in the `strained sample'.
A controllable strain engineering should be explored in the future to tune the
electronic phases of 1\emph{T}-TaS$_2$.

\noindent\textbf{Methods}

\noindent\textbf{Sample preparation.} The high quality 1\emph{T}-TaS$_2$ single crystals were grown
by the chemical vapor transport (CVT) method with iodine as the transport agent.
Ta (99.99$\%$, Aladdin) and S (99.99$\%$, Aladdin) powders with mole ratio of 1:2 were weighted
and mixed with 0.2 g of I$_2$, which were placed into silicon quartz tubes.
These tubes were sealed under high vacuum and heated for 10 days in a two-zone furnace,
where the temperature of source and growth zones were fixed at 850 $^\circ$C and 750 $^\circ$C, respectively.
Then the quartz tubes were removed from the furnace and quenched in ice water mixture.

\noindent\textbf{STM measurement.} The STM and STS experiments were carried out in a
commercial STM system~\cite{ZhengSREP17}. The samples were gradually cooled to liquid
nitrogen temperature and in situ cleaved. An electrochemically etched tungsten
tip was treated with e-beam sputtering and field emission on a single crystalline with
an Au (111) surface. A constant current scanning mode maintained by a feedback loop
control was used in this experiment. The $dI/dV$ spectra were taken with a standard
lock-in technique with a frequency of 983.4 Hz and an amplitude of 10 mV. The $dI/dV$
spectra at domain walls were taken with a bias modulation of 2 mV. All data were
acquired at liquid helium temperature ($\sim$~4.5 K) except for special statement.

\noindent\textbf{Data availability.} The source data and related supporting information are available upon
reasonable request from the corresponding author.

\noindent\textbf{Acknowledgments}

\noindent{This work was supported by the National Basic Research Program of China (Grant No. 2015CB921004),
the National Natural Science Foundation of China (Grant No. NSFC-11374260),
and the Fundamental Research Funds for the Central Universities in China.
J.J.G., X.L. and Y.P.S. thank the support of the National Key Research and Development Program under contracts 2016YFA0300404,
the National Nature Science Foundation of China under contracts 11674326 and 11874357,
the Joint Funds of the National Natural Science Foundation of China,
and the Chinese Academy of Sciences' Large-Scale Scientific Facility under contracts U1832141.}

\noindent\textbf{Author contributions}

\noindent{K.L.B. and W.H.Z. conducted the STM experiment. K.L.B. analyzed the data. Z.X.W. participated in the experiment.
Y.F. and Y.Z. discussed the results. J.J.G., X.L. and Y.P.S. grew the samples.
K.L.B. and Y.Y. wrote the paper. Y.Y. supervised the experiment.
All authors have read the paper and approved it.}

\noindent\textbf{Additional information}

\noindent\textbf{Competing financial interests:} The authors declare no competing financial or non-financial interests.

\end{document}